%% file: sameen_2006.tex
\documentclass[referee]{jfm}

\usepackage{graphicx}
\usepackage{natbib}

\ifCUPmtlplainloaded \else
  \checkfont{eurm10}
  \iffontfound
    \IfFileExists{upmath.sty}
      {\typeout{^^JFound AMS Euler Roman fonts on the system,
                   using the 'upmath' package.^^J}%
       \usepackage{upmath}}
      {\typeout{^^JFound AMS Euler Roman fonts on the system, but you
                   dont seem to have the}%
       \typeout{'upmath' package installed. JFM.cls can take advantage
                 of these fonts,^^Jif you use 'upmath' package.^^J}%
      }
  \else
  \fi
\fi

\ifCUPmtlplainloaded \else
  \checkfont{msam10}
  \iffontfound
    \IfFileExists{amssymb.sty}
      {\typeout{^^JFound AMS Symbol fonts on the system, using the
                'amssymb' package.^^J}%
       \usepackage{amssymb}%
       \let\le=\leqslant  
         
      }{}
  \fi
\fi

\ifCUPmtlplainloaded \else
  \IfFileExists{amsbsy.sty}
    {\typeout{^^JFound the 'amsbsy' package on the system, using it.^^J}%
     \usepackage{amsbsy}}
    {\providecommand\boldsymbol[1]{\mbox{\boldmath $##1$}}}
\fi

\newsavebox{\astrutbox}
\sbox{\astrutbox}{\rule[-5pt]{0pt}{20pt}}

\newcommand\p{\ensuremath{\partial}}

\usepackage{defn}
\usepackage{multirow}
\usepackage{subfig}

\title[Stability of heated channel flow]{The effect of wall heating
on instability of channel flow}

\author[A. Sameen and Rama Govindarajan]%
 {A.\ns S\ls A\ls M\ls E\ls E\ls N%
\ns
\and R\ls A\ls M\ls A\ls ~G\ls O\ls V\ls I\ls N\ls D\ls A\ls R\ls A\ls 
J\ls A\ls N}

\affiliation{
Engineering Mechanics Unit, Jawaharlal Nehru Centre for Advanced 
Scientific Research, Bangalore-560064, INDIA
}

\pubyear{1996}
\volume{538}
\pagerange{119--126}
\date{?? and in revised form ??}

\begin{document}

\maketitle

\begin{abstract}
A comprehensive study of the effect of wall heating or cooling on the 
linear, transient and secondary growth of instability in channel flow
is conducted. The effect of viscosity stratification, heat diffusivity 
and of buoyancy are estimated separately, with some unexpected results.
>From linear stability results, it has been accepted that heat diffusivity 
does not affect stability. However, we show that realistic Prandtl 
numbers cause a transient growth of disturbances that is an order of 
magnitude higher than at zero Prandtl number.
Buoyancy, even at fairly low levels, gives rise to high levels of 
subcritical energy growth. Unusually for transient growth, both of these 
are spanwise-independent and not in the form of streamwise vortices. 
At moderate Grashof numbers, exponential growth dominates, with distinct
Rayleigh-Benard and Poiseuille modes for Grashof numbers upto $\sim 25000$,
which merge thereafter.
Wall heating has a converse effect on the secondary instability compared
to the primary, destabilising significantly when viscosity decreases
towards the wall. It is hoped that the work will motivate experimental
and numerical efforts to understand the role of wall heating in the control
of channel and pipe flows.
\end{abstract}

\input{intro}

\input{mean}

\input{linear}
\input{transient}

\input{prgr}

\input{secondary}
\input{conc}


\bibliographystyle{jfm}

\end{document}

%% file: intro.tex
\section{Introduction}

One of the well-known methods for delaying a transition to turbulence,
for example in boundary layers, has been to reduce the viscosity at the 
wall. Such a reduction could be brought about by heating or cooling the
surface, for example. The objective of this paper is to study the effect
of wall heating on the instability of a channel flow. It is shown that
heat can have surprising effects on the different mechanisms of
transition. We restrict ourselves here to routes based on the linear 
eigenmodes, a direct nonlinear interaction will be studied in future. 
The emphasis here is on delaying/advancing the onset of transition
to turbulence, rather than drag reduction in full turbulence, as
achieved by adding small quantities of polymer. 
 
The critical Reynolds number for linear instability in a plane Poiseuille 
flow is $5772.22$ [\cite{orszag71}]. However, experiments usually
find fully developed turbulence at a much lower Reynolds number, 
around $1500$ [see e.g. \cite{davies,narayanan,patel,kao}].
It is clear that routes to turbulence other than the traditional
Tollmien-Schlichting (TS) mechanism are in operation. 
The background noise in the flow has a
major influence in delaying/hastening transition to turbulence, as
well as in deciding which mechanism will be dominant [\cite{morko89}]. 
At extremely low
levels of noise a traditional TS mechanism and/or secondary 
instability is likely to be followed. At intermediate levels,
a transient growth of disturbances is the more likely mechanism 
for initial disturbance growth [\cite{schmid,alv,foster,corbet}]. 
Once disturbance growth is triggered by a linear mechanism, 
nonlinearities are required to achieve a new self-sustained state.
Alternatively, at higher levels of background noise, nonlinear mechanisms
can directly come into play, see e.g. \cite{wal1,holg03,hof}.
At present it is not understood exactly which route will be followed 
when (for a recent review on pipe flow see \cite{kerswell2005}). 

The effect of wall heating on {\em linear stability} alone has been 
studied by several researchers. Here too, the effect of buoyancy
has not been clearly quantified. To our knowledge, a detailed
study of other mechanisms has not been done. 
Two related studies of transient growth had different emphasis from 
the present work.
Transient growth in two-fluid flow was studied in two-dimensions
by \cite{alison} 
with the objective of understanding the effect of the interface.
\cite{bott2004} studied transient growth with
stable thermal stratification and concluded that such
stratification is a viable strategy to control transitional flows.
A more detailed retrospective on earlier work is included in the relevant 
sections later in the text.

We consider two types of heating. The first is asymmetric, with
the two walls maintained at different constant temperatures. The second
is symmetric, with the walls at one temperature and the fluid at another.
Our results may be summarised as follows: the linear stability
results are in line with the findings of earlier studies. A 
decrease in viscosity as one approaches a wall has a large stabilising 
effect and vice-versa. The effect on the linear eigenmode of reduced 
heat diffusion (increasing Prandtl number) is extremely small 
[\cite{wall}]. Buoyancy has no effect up to a Grashof number
of about $3000$ and is enormously stabilising or destabilising thereafter, 
depending on the sign of the temperature difference. The Rayleigh-Benard
and Tollmien-Schlichting modes are distinct at low Grashof number
and merge at high Grashof numbers.

The effect of heat on transient growth of instabilities is unexpected.
Viscosity stratification, which is the chief player in linear
instability, has no discernible effect on this mechanism. Increasing
Prandtl number, on the other hand, has an order of magnitude destabilising
effect. The assumption that Prandtl number may be neglected in stability
analyses is therefore completely incorrect for this mechanism.
Secondary instabilities of the Tollmien-Schlichting modes are usually
taken to be unimportant for channel-flow transition, but we find that
viscosity-stratification can have a destabilising effect on these
modes, which may make them noticeable at large temperature differences.

%% file: mean.tex
\section{Basic Velocity Profiles}
\label{velo_sec}

\begin{figure}
\centering
\begin{minipage}{0.45\textwidth}
\includegraphics[width=1.0\textwidth]{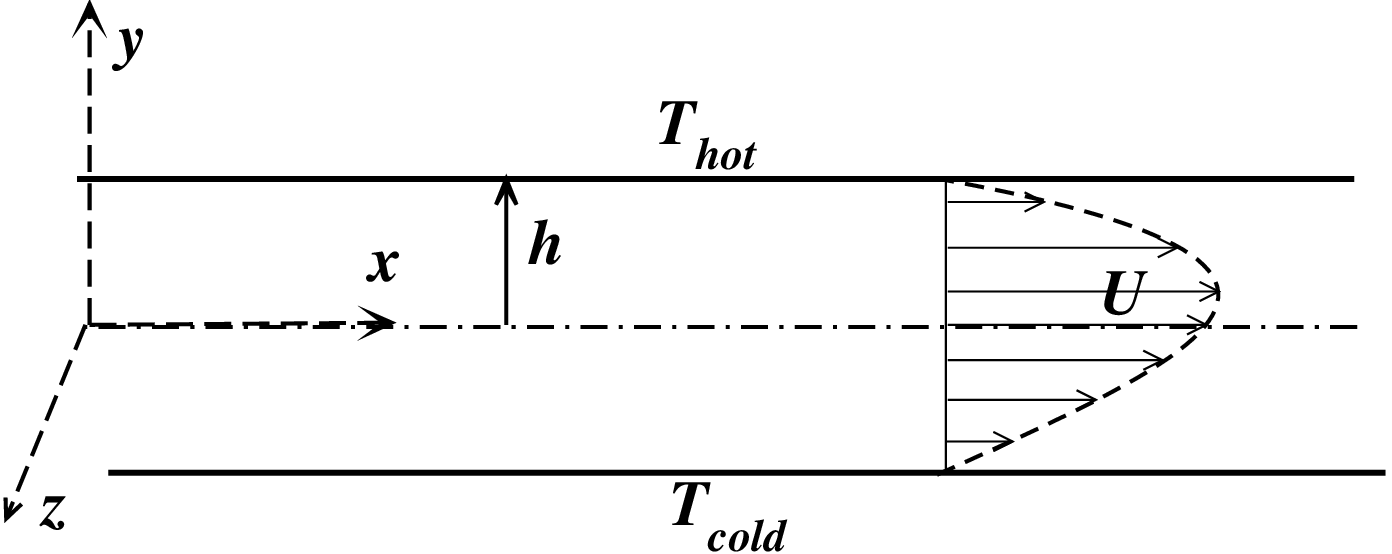}
\caption{Schematic diagram of the channel, asymmetric heating.
The two walls are held at different temperatures,
$T_{hot}$ and $T_{cold}$, so the mean temperature profile is linear. }
\label{chan}
\end{minipage}
\hskip5mm
\begin{minipage}{0.45\textwidth}
\includegraphics[width=1.0\textwidth]{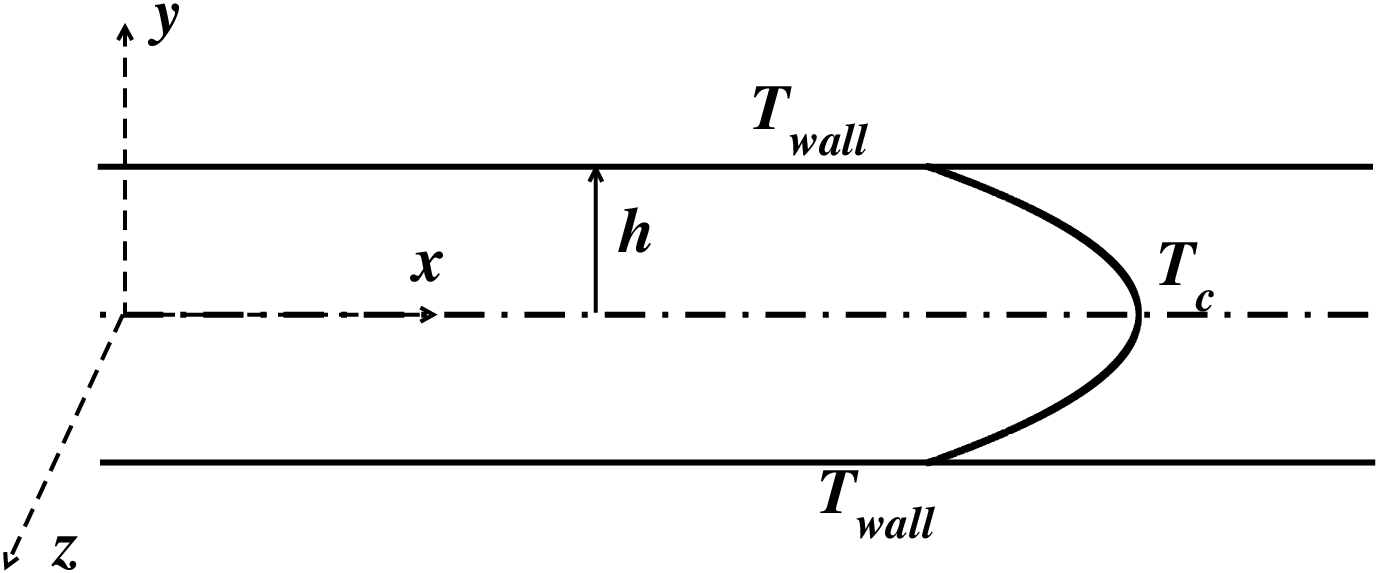}
\caption{Schematic diagram of the channel, symmetric heating. Both walls
are maintained at the same temperature, different from that of the 
fluid.}
\label{cha5th}
\end{minipage}
\end{figure}

Two types of temperature variation, which we shall refer to as the
asymmetrically and symmetrically heated cases respectively, are 
considered, see figures \ref{chan} and \ref{cha5th}. These provide a 
fair sample of the type of stratification we may come across.
In the first, the two walls of the channel are maintained at different 
temperatures, $T_{hot}$ and $T_{cold}$. At steady state, the 
temperature within the channel varies linearly between the two.
Note that for the unstable Poiseuille-Rayleigh-Benard configuration, 
the temperature difference $\Delta T$ between the bottom and top
walls
(and hence the corresponding Richardson and Grashof numbers, defined 
later) is taken to be positive. The sign of $\Delta T$ is unimportant
when buoyancy is neglected.

\begin{figure}
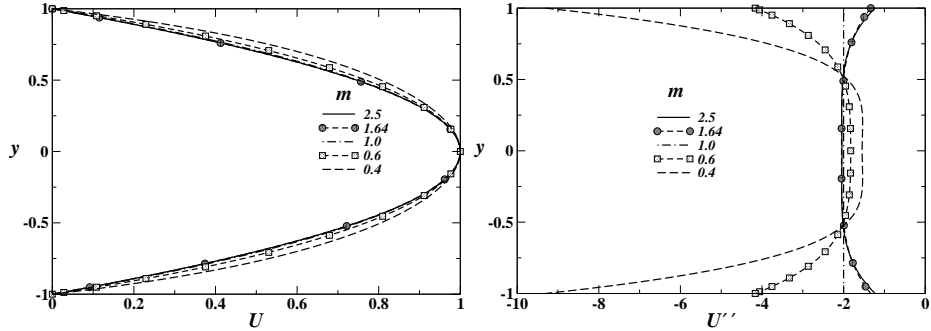

\centering
\begin{minipage}{1.0\textwidth}
\includegraphics[width=0.45\textwidth]{figs/u.eps}
\includegraphics[width=0.45\textwidth]{figs/udd.eps}
\caption{ Variation of non-dimensional velocity and
its second derivative $U^{\prime\prime}$ with symmetric heating.}
\label{u_sym}
\mine
\end{figure}

In the second case, the walls are both maintained at the same 
temperature, while the incoming fluid is at a different temperature. 
Such a flow is non-parallel, since the fluid temperature downstream 
tends to equilibrate with the wall temperature, but for large Peclet 
numbers, the change in the downstream direction is very slow, and
the flow may safely be assumed to be locally parallel and the local
temperature profile to be parabolic. 
Stability analyses for more realistic temperature profiles have been 
conducted, without any qualitative difference. 
The temperature-dependence of the viscosity is described by the
Arrhenius model, which works fairly well for most common liquids like 
water and alcohol.
\be
\mu(T)=C_1 \exp(C_2/T),
\label{visc_eqn}
\ee
where $C_1$ and $C_2$ are constants associated with the fluid under 
consideration, which is taken in the present computations to be water.
The streamwise direction is denoted as $x$, the coordinate $y$ is 
normal to the wall, and  $z$ is the spanwise direction.
The mean $x$-momentum equation for a plane parallel channel flow 
reduces to
\begin{equation}
(\mu U^\prime)^\prime=\ddpopa{P}{x} Re,
\label{basic}
\end{equation}
where the primes denote differentiation with respect to $y$,
$Re$ is the Reynolds number defined as $Re={U_{max}h}/{\nu_{ref}}$,
$h$ is the half-channel width, $\nu_{ref}$ is the reference kinematic 
viscosity, and $\ddpopas{P}{x}$ is the constant pressure gradient.
The viscosity 
ratio is defined as $m=\mu_{cold}/\mu_{hot}$ for the asymmetric
heated case and $m=\mu_{wall}/\mu_{c}$, 
where $c$ stands for centerline, for the symmetric case.
Knowing $\mu(T)$ and $T(y)$, equation \ref{basic} is integrated twice 
by a fourth-order
Runge-Kutta method to get $U$. 
Figure \ref{visc_asym} shows typical viscosity and velocity profiles 
for asymmetric heating. Viscosity profiles for the symmetric case are 
shown in figure \ref{mu_sym}.
\begin{figure}
\centering
\minps
\includegraphics[width=1.0\textwidth]{figs/vel_visc.eps}
\caption
{Variation of (a) viscosity and (b) velocity with asymmetric 
heating. The velocity is scaled by its maximum and the viscosity is 
scaled here by its value at the hot wall.}
\label{visc_asym}
\mine
\hskip8mm
\minps
\includegraphics[width=1.0\textwidth]{figs/mu.eps}
\caption{Variation of viscosity, scaled here by its value at the 
centerline, for symmetric heating.}
\label{mu_sym}
\mine
\end{figure}
Corresponding velocity profiles and their second derivatives are 
shown in figure \ref{u_sym}.

\begin{center}
\begin{table*}
\begin{minipage}{180mm}
\hskip20mm
\begin{tabular}{cccccccccc} \hline
& $\Delta T ^0K$&&&$m$ &&& $\mu_{avg}$ &&  $U_{avg}/U_{avg,m=1.0}$\\
\hline
&$50$&&&0.39&&&0.748&&1.068\\
&$10$&&&0.81&&&0.933&&1.006\\
&$0.0$&&&1.00&&&1.000&&1.000\\
&$-10$&&&1.23&&&1.070&&0.995\\
&$-50$&&&2.51&&&1.381&&0.9944\\ \hline
\end{tabular}
\end{minipage}
\caption{The dependence of the average viscosity $\mu_{avg}$ on
the viscosity ratio, $m$,
symmetric heating. The last column shows the ratio of the average
mean velocity to its value in the unheated case. The ratio is close to
$1.0$ in all cases. }
\label{reavg} \vskip4mm
\end{table*}
\end{center}

Unless otherwise specified, we define the Reynolds number in terms of the 
average viscosity across the channel, as follows.
\beqa
Re \equiv \frac{U_{max}h}{\int_{-1}^{0}\mu dy}. 
\eeqa
We can see from Table \ref{reavg} that the average viscosity varies
significantly with increasing $\Delta T$, so defining the Reynolds number as
above is approriate for making comparisons at a given Reynolds number between
heated and unheated flows. On the other hand, it is seen from the table that
the average velocity is practically unchanged by viscosity stratification, so
the maximum velocity is a good enough velocity scale.

%% file: linear.tex
\section{Linear stability}\label{sec:prim}
\label{3d_sec}

\subsection{The stability equations and their solution}

The disturbance quantities in normal mode form are given as
\begin{equation}
\Big[\hat v,\hat \eta, \hat T\Big] =\Big[v(y),\eta(y),\hat T(y)\Big]
\exp[i(\alpha x +\beta z-\omega t)],
\label{uvwp}
\end{equation}
where $\hat v$ and $\hat \eta$ respectively are the components of
disturbance velocity and vorticity in the direction normal to the wall, $\hat T$ is
the disturbance temperature, $\alpha$ 
and $\beta$ are the wave numbers in the streamwise and spanwise directions 
respectively,
and $\omega$ is the complex frequency of the wave.
The linear stability equations may be derived to be
\begin{eqnarray}
&&i\alpha\left[(v^{\prime\prime}-(\alpha^2+\beta^2))(U-c)-U^{\prime\prime}
v\right]
=\frac{1}{Re}\Big[\mu \left[v^{iv}-2(\alpha^2+\beta^2)v^{\prime\prime}
+(\alpha^2+\beta^2)^2 v\right]
\nonumber\\
&&+\ddpopa{\mu}{T}T^{\prime}2[v^{\prime\prime\prime}-(\alpha^2+\beta^2)v^{\prime}]
+\ddpopa{\mu}{T}T^{\prime\prime}[v^{\prime\prime}+(\alpha^2+\beta^2) v]
+\dpopasq{\mu}{T}T^{\prime\prime}[v^{\prime\prime}+(\alpha^2+\beta^2) v]
\nonumber \\ &&
+\ddpopa{\mu}{T}[U^{\prime}\hat{T}^{\prime\prime}+2U^{\prime\prime}\hat{T}^{\prime}
+(\alpha^2U^{\prime}+U^{\prime\prime\prime})\hat{T}]
+2\dpopasq{\mu}{T}U^{\prime}T^{\prime}\hat{T}^{\prime}
+\dpopasq{\mu}{T}T^{\prime\prime}U^{\prime}\hat{T}
\nonumber\\ &&
+\dpopacb{\mu}{T}U^{\prime}T^{\prime}\hat{T} {-\frac{Gr}{Re} i \alpha \hat{T}}\Big],
\label{orr_eqn}
\end{eqnarray}
\begin{eqnarray}
&&\alpha (c-U) \eta +i\beta U^{\prime}v=\frac{1}{Re}\Big[\mu \big[\eta^{\prime\prime}
-(\alpha^2+\beta^2)\eta\big] +\ddpopa{\mu}{T}T^{\prime}\eta^{\prime}
-i\beta\ddpopa{\mu}{T}(U^{\prime\prime}\hat{T} +U^{\prime}\hat{T}^{\prime})
\nonumber \\ &&
-i\dpopasq{\mu}{T}T^{\prime}U^{\prime}\hat{T}
\Big],
\label{sq_eqn}
\end{eqnarray}
\begin{eqnarray}
i\alpha \Big [(U-c)\hat T-{T}^{\prime}v\Big ]
=\frac{1}{RePr}\Big [{\hat T}^{\prime\prime}-(\alpha^2 + \beta^2)\hat T\Big ],
\label{eneT_eqn}
\end{eqnarray}
where $c\equiv\omega/\alpha$.
Equations \ref{orr_eqn} and \ref{sq_eqn} respectively are the
Orr-Sommerfeld and Squires equations, modified here to account for
the effects of viscosity variations, temperature fluctuations and
of buoyancy. A Boussinesq approximation has been made. If buoyancy
were to be neglected, the above equations would be equivalent to 
those of Wall \& Wilson (1996).

The Prandtl number is defined as $Pr \equiv {\nu/\kappa}$
where $\kappa$ is the coefficient of 
thermal diffusivity.
The Grashof number is $Gr \equiv {g\gamma \Delta T h^3}/
{\nu^2}$, $\gamma$ being the volume coefficient of expansion.
Equations \ref{orr_eqn} to \ref{eneT_eqn} form an eigenvalue problem 
with the boundary conditions
\begin{equation}
v(\pm1)=v^\prime(\pm1)=\eta(\pm1)= \hat T (\pm1) = 0,
\label{bc3d}
\end{equation}
and are solved using a Chebyshev collocation spectral method.
We perform a temporal stability analysis, the growth rate of the
disturbance is obtained from the imaginary part of $c$.

\subsection{Effect of viscosity variation}

In order to isolate the effect of viscosity variation, the Prandtl
number and the Grashof number are set to zero. With symmetric
heating, we expect that if the viscosity decreases as one approaches
the wall, the fuller velocity profile will result in a stabler flow.
This expectation is realised, as seen from the neutral stability 
boundaries in figure \ref{neu_case2}.

However, in a channel
flow where one wall is maintained at a constant high temperature
and the other wall is kept cold, the viscosity
decreases towards one wall and increases towards the other. It is
not a priori evident what the effect on the linear stability will
be. 
It was found by {\cite{potter}} that any temperature difference 
between the walls is always destabilising. However,
{\cite{wall}} found, using four different viscosity models, that a 
temperature difference almost always stabilises the flow.
The apparent contradiction is because the former work
compared results for heated and unheated flow maintaining the input 
power constant, while the latter made comparisons at a given Reynolds
number. Since the flow rate for a given input
power is higher for the heated case, the resulting Reynolds number is 
higher. 
The stability of viscosity-stratified channel flows was also studied 
by \cite{pina,sch93} with similar conclusions and of boundary layer 
flows by \cite{craik71,kao68,wazzan68,wazzan79,resh77,sch95}. 

\begin{figure}
\centering
\minp
\includegraphics[width=0.8\textwidth]{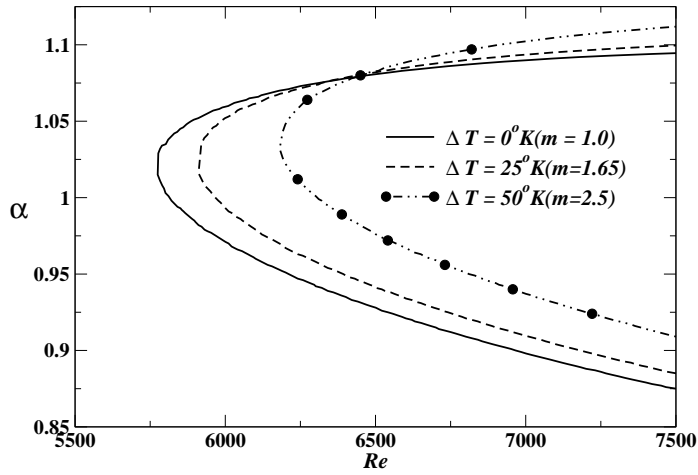}
\caption {Stability boundaries for various viscosity ratios. Asymmetric
heating, $T_{cold}=295^oK$.} 
\label{neu_aysm_fig}
\mine
\end{figure}

In the present paper, we define the Reynolds number in terms
of average viscosity, and compare results at a given Reynolds number.
In agreement with {\cite{wall}}, for asymmetric heating,
we find that any temperature difference
is stabilising, in terms of the least stable (two-dimensional) linear mode
(figure \ref{neu_aysm_fig}). We have confirmed [\cite{phd}] that 
the production of disturbance kinetic energy is reduced at the cold wall
and increased at the hot wall compared to the unheated case. The
dissipation is similar in all cases.
The highly oblique modes, unlike the two-dimensional ones,
are practically unaffected (not shown). This 
observation will assume significance when we discuss transient growth.

\begin{figure}
\centering
\minp
\includegraphics[width=0.8\textwidth]{figs/rcrit_symmT.eps}
\caption
{Effect of viscosity variation on linear stability, symmetric heating.
For unstratified flow, i.e., at $m=1.0$, $Re_{cr}$ is 5772.2.} 
\label{neu_case2}
\mine
\end{figure}

\subsection{Effect of heat diffusivity}

\begin{figure}
\centering
\minp
\includegraphics[width=0.8\textwidth]{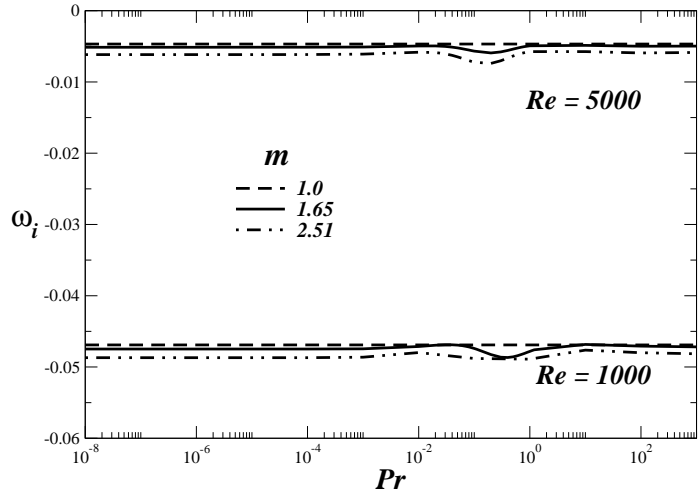}
\caption {Most unstable eigenvalue at various Prandtl numbers for 
different $\Delta T$ at $\alpha=0.9$ and $Re=1000,5000$. The effect
of Prandtl number is negligible.} 
\label{eig_pec_fig}
\mine
\end{figure}

We know that for liquids such as water, heat diffuses slower than 
momentum, so the
assumption of $Pr=0$ is not justifiable. Surprisingly however, the 
linear stability, as measured by the least stable eigenmode, is practically
unaffected by a decrease in heat diffusivity [\cite{wall}]. Present
computations confirm this (figure \ref{eig_pec_fig}).
However, the prevalent conclusion that heat diffusivity does not 
affect flow stability, and therefore that the Peclet number may be
set to zero in stability analyses, is shown in the next section to be 
incorrect. Increasing the Prandtl number to $O(1)$ vaules
can enhance transient growth by an order of magnitude. 

\subsection{Effect of buoyancy: the Poiseuille-Rayleigh-Benard problem}
\label{sec:linbuoy}
We consider the asymmetrically heated case here. When the upper
wall is cold relative to the lower one, the resulting unstable
stratification of density leads, at low flow rates, to a
buoyancy driven instability similar to the Rayleigh-Benard 
[\cite{chandra61,turner69,platten}].
The effect of mean shear on this instability has been studied by
\cite{lib1,dear65}, for example, and this problem is reviewed in
\cite{platten,gebh}. On the other hand, the effect of buoyancy on 
the Tollmien-Schlichting modes in plane
Poiseuille flow has been investigated by \cite{gage68,gage71,
tveit74} and \cite{fujimura}. Several approximations were made in
these early studies. For example, viscosity variations were neglected
and the base flow was taken to be parabolic. 
It was found
that a critical Reynolds number always exists for any level of
density stabilisation, while there exists a Richardson number 
($Ri\equiv Gr/Re^2$) above which the flow is stable for all Reynolds 
number. 



\begin{figure}
\centering
\minp
\includegraphics[width=0.8\textwidth]{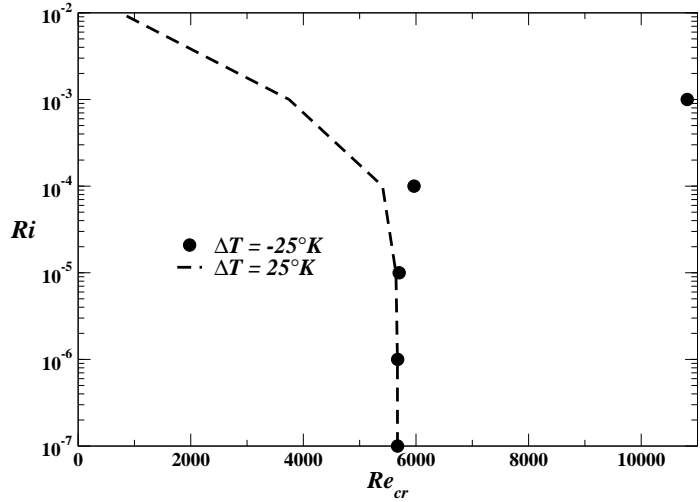}
\caption
{Neutral stability Reynolds number as a function of Richardson number,
$Pr=1.0$. Note that for stable stratification ($\Delta T<0$), the
Richardson number is negative, its absolute value is plotted here. The portion
to the right of the neutral points is unstable. }
\label{neu_uns}
\mine
\end{figure}

\begin{figure}
\centering
\minp
\includegraphics[width=0.8\textwidth]{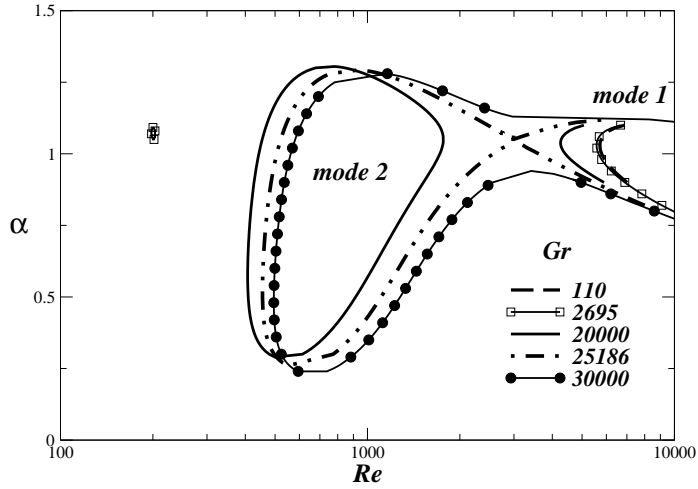}
\caption
{The neutral curves for unstably thermal stratified flow at Prandtl 
number $7.0$, $\Delta T=25^oK$. The second (Rayleigh-Benard-like) mode 
starts appearing at $Gr=2695$ and merges with the 
Tollmien-Schlichting mode at $Gr=25186$. }
\label{grneu}
\mine
\end{figure}

\begin{figure}
\centering
\minp
\includegraphics[width=0.8\textwidth]{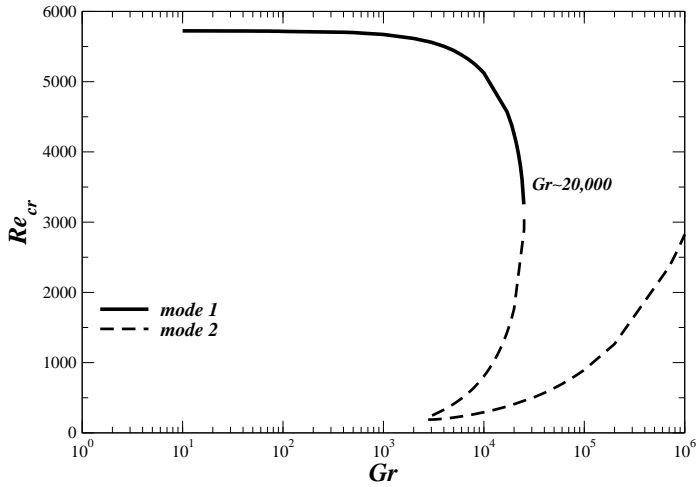}
\caption
{A consolidated picture of the variation of the critical Reynolds
number with the Grashof number. The Tollmien-Schlichting mode
is shown by the solid line, and the Rayleigh-Benard mode by the
dashed line. The region above the curves is unstable.}
\label{grdom}
\mine
\end{figure}

In figure \ref{neu_uns} the critical Reynolds number,$Re_{cr}$ for a 
temperature difference of $25^oK$ is plotted 
for various Richardson numbers. The trends are the same as in 
\cite{gage68} and \cite{tveit74}, but there are minor numerical 
discrepancies, which we attribute to the more appropriate velocity 
and viscosity profiles used here. The effect of buoyancy are negligible
when the Richardson number is below $10^{-4}$, and of either sign.
At higher Richardson numbers, for unstable stratification,
figure \ref{neu_uns} shows that the flow is highly destabilised.
The stability boundaries are plotted in figure \ref{grneu} in terms
of the Grashof number, a given Grashof number being more simple
to achieve experimentally. 
Distinct modes of Rayleigh-Benard type and of Tollmien-Schlichting
type are evident at intermediate levels of $Gr$. The modes merge
at Grashof numbers above $\sim 25000$. The numerical value at the bifurcation
point varies slightly with Prandtl number and temperature difference. The 
unstable region in the Grashof-Reynolds parameter space is shown in 
figure \ref{grdom}.

We have discussed the stability in terms of the most
unstable linear mode. However, a transient growth of decaying
modes can often be the dominant mechanism of transition to 
turbulence in channel flows. We shall see in the next section
that the effect of heat on flow instability throws up
several surprises.

%% file: transient.tex
\section{Transient Growth}
 
The linear stability operator is not
self-adjoint, and the resulting non-orthogonality of the
eigenfunctions is known to be able to give rise to large levels of
transient growth of disturbance kinetic energy even when all individual
eigenmodes are stable. In wall-bounded flows, transient growth is 
mainly caused by the interaction between the Orr-Sommerfeld and 
Squire modes [\cite{red93,crim03}] from the coupling term,
$-i\beta U^{\prime}$, appearing in Squire's equation.
The most likely structures arising due to transient growth are
streamwise streaks [\cite{hen98,SH,red93,schmid}].
We use the standard approach for computing the maximum transient 
growth.

The effect of viscosity-stratification, in contexts other than
heat [\cite{alison,vijay}] has been addressed before, though not
completely. The effect of buoyancy has been studied under stable
stratification alone by \cite{bott2004}. The effect of heat diffusivity on 
transient growth has not been studied before, to our knowledge. 

\begin{figure}
\centering
\minp
\includegraphics[width=0.8\textwidth]{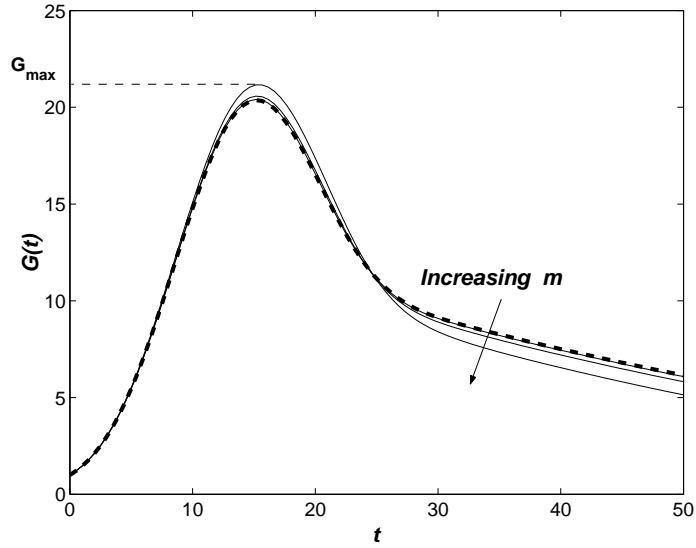}
\caption{The energy amplification evolution for various viscosity 
ratios for $Re=3000, \alpha=1$, asymmetric heating. The thick dashed curve
is for unstratified flow, and the solid lines are for $m=1.2, 1.65$ and $2.5$
in the order indicated in the figure. }
\label{gt2}
\mine
\end{figure}

The disturbance kinetic energy, $g(t)$ [\cite{schmid}], is written as
\beq
g(t)=\frac{\|\kappa(t)\|^2_E}{\|\kappa(0)\|^2_E}=
\frac{\|e^{-i\Lambda t} \kappa(0)\|^2_E}{\|\kappa(0)\|^2_E}.
\label{gt}
\eeq
Its time 
evolution is represented by the matrix $\p{\kappa}{t}=-i\Lambda \kappa$,
where $\kappa=(\kappa_1,\kappa_2,...,\kappa_N)^T$
and $\Lambda=diag\{\omega_1,\omega_2,...\omega_N\}$,
where $\kappa_j$ is the $j^{th}$ expansion coefficient of the
eigenfunctions of the linear modes which are the dominant contributors
here. The superscript $T$ denotes transpose.
Maximising equation (\ref{gt}) for all possible initial conditions $\kappa(0)$,
\begin{eqnarray}
G(t)=\max_{\kappa \ne 0} g(t).
\label{gtmax}
\end{eqnarray}
We then define $G_{max}$ as the maximum over time of $G(t)$ for one 
particular $Re,\alpha$ and $\Delta T$, see figure \ref{gt2}.

\begin{figure}
\centering
\minp
\includegraphics[width=0.8\textwidth]{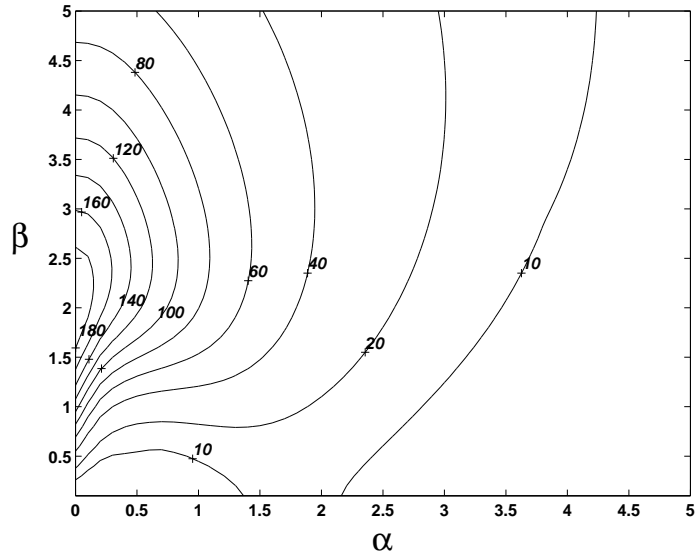}
\caption{The contour of $G_{max}$ (the maximum over time of $G(t)$) for 
$Re=1000$ in the $\alpha-\beta$ plane, $\Delta T=0^oK$. This matches well with \cite{red93}}
\label{gmax00}
\mine
\end{figure}
The contour plot for $G_{max}$ for unheated flow is shown 
at $Re=1000$ in figure \ref{gmax00}, for comparison with the results for 
heated flow to follow. A maximum growth of $G_{max}=196$ is obtained 
for $\alpha=0.0$ and $\beta=2.05$ [see \cite{red93}].

\subsection{Effect of viscosity stratification}
\begin{figure}
\centering
\minp
\includegraphics[width=0.8\textwidth]{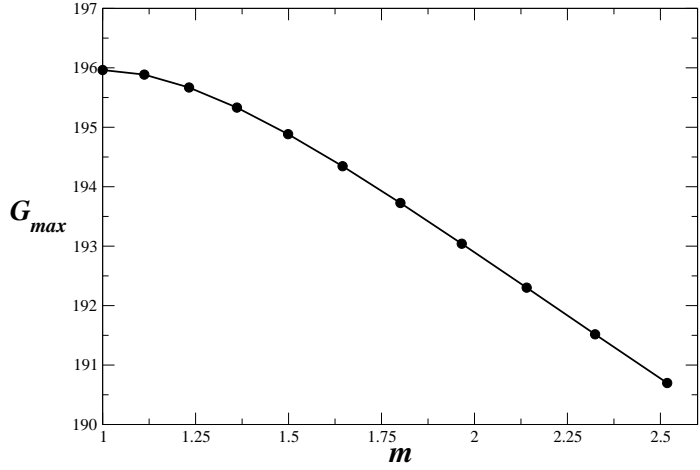}
\caption
{The variation of $G_{max}$ at $\alpha=0$ and $\beta=2$ for various
viscosity ratios at $Re=1000$, asymmetric heating. The maximum deviation of $G_{max}$ 
from the unheated value of $196$ is only $3\%$.}
\label{gmax_m}
\mine
\end{figure}

\begin{figure}
\centering
\minp
\includegraphics[width=0.8\textwidth]{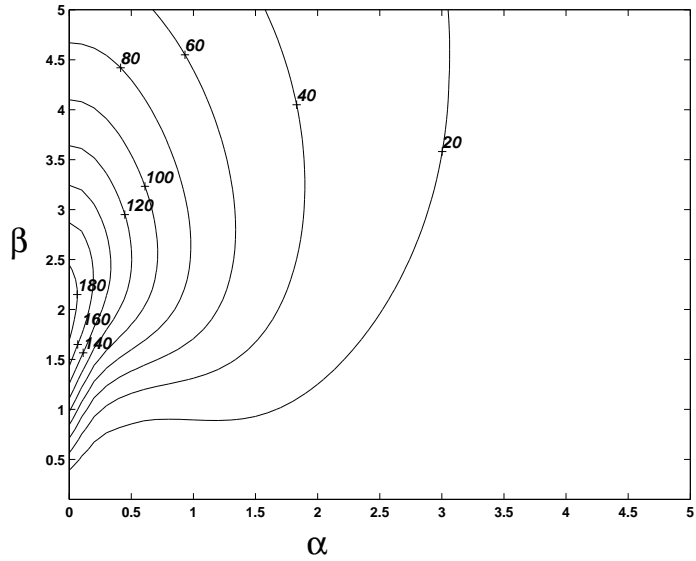}
\caption{The contour of $G_{max}$ for symmetric heating at $Re=1000$ again
in the $\alpha-\beta$ plane, $m=0.4$.}
\label{gmaxm50ml}
\mine
\end{figure}

As before, we first take the Prandtl number to be zero, i.e., assume
that temperature
fluctuations diffuse away instantaneously. We also neglect buoyancy, in
order to isolate the effect of viscosity stratification alone.
For the asymmetrically heated case, the growth of kinetic energy 
is seen in figure \ref{gt2} to change only marginally 
with heating. 
The example shown in figure \ref{gt2} is for $Re=3000$
and $\alpha=1$, but viscosity stratification has very little effect
on transient growth at any value of $Re$ and $\alpha$.
The effect of asymmetric heating is quantified in
figure \ref{gmax_m} in terms of $G_{max}$ at $\alpha=0$ and $\beta=2$. 
There is a marginal 
stabilisation with viscosity stratification. This result is in line with the
result for linear stability, but much smaller in magnitude.
The contours of $G_{max}$ for symmetric heating are plotted in 
figures \ref{gmaxm50ml} in the $\alpha-\beta$ plane, and
\begin{figure}
\centering
\minp
\includegraphics[width=0.8\textwidth]{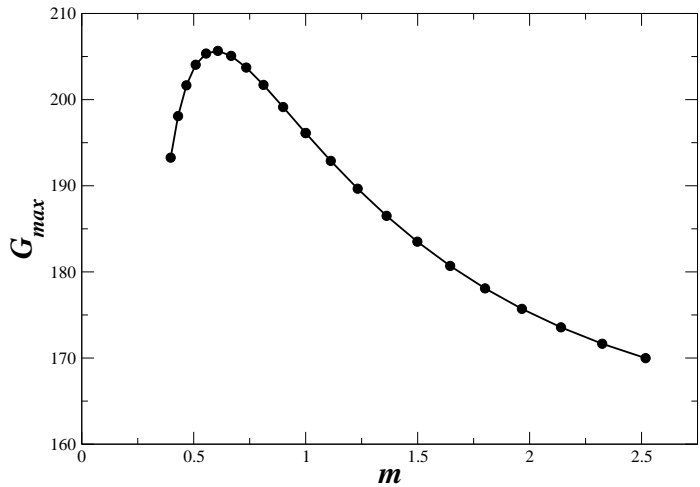}
\caption
{ The $G_{max}$ variation at $\alpha=0.0$ and $\beta=2.0$ for
various viscosity ratios at $Re=1000$, symmetric heating. The maximum deviation 
of $G_{max}$ from that for unstratified flow is only $13\%$.}
\label{gmax_symm}
\mine
\end{figure}
the variation of $G_{max}$ with viscosity ratio is plotted in
figure \ref{gmax_symm} at $\alpha=0$. There is again a slight 
stabilisation with increase in viscosity stratification.

\begin{figure}
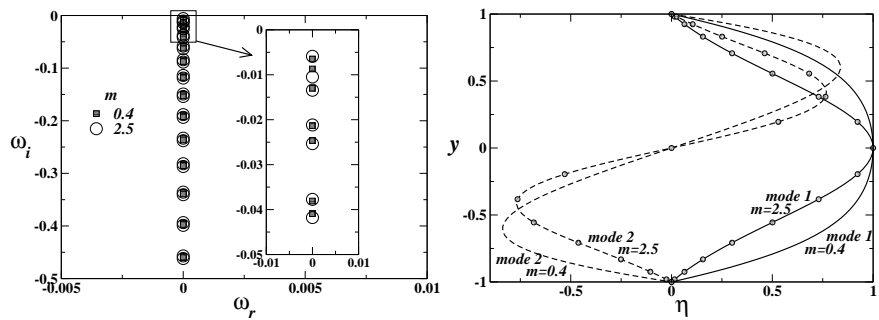

\centering
\minp
\includegraphics[width=0.5\textwidth]{figs/eigval_re1000_beta2_al0.eps}
\includegraphics[width=0.5\textwidth]{figs/eig_eta_re1000_al0_beta2.eps}
\caption
{(a) The eigenspectra for two extremes of viscosity stratification, $m=0.4$
and $2.5$, for $\alpha=0$, $Re=1000$ and $\beta=2.0$, symmetric heating.
(b) The corresponding eigenfunctions of the first two unstable eigenvalues in
each case. 
}\label{spec_eigal0}
\mine
\end{figure}

The insignificant effect of viscosity stratification is consistent with 
our recent study of transient growth in two-fluid and non-Newtonian 
flows [\cite{vijay}]. As discussed there, the $U^{\prime\prime}$ term,
which affects the least stable eigenmode dramatically, has no effect
on streamwise vortices arising from $\alpha=0$, which dictate transient
growth. The eigenspectrum, and typical eigenfunctions at $\alpha=0$ are
shown in figure 
\ref{spec_eigal0} to be very similar at two extremes of viscosity 
stratification.
Equation \ref{sq_eqn} drives the dynamics rather than equation 
\ref{orr_eqn} under these conditions, and the terms
containing viscosity gradients have been verified
numerically to be small.

%% file: prgr.tex
\subsection{Effect of heat diffusivity}
\label{sec:pec}
\begin{figure}
\centering
\minp
\includegraphics[width=0.9\textwidth]{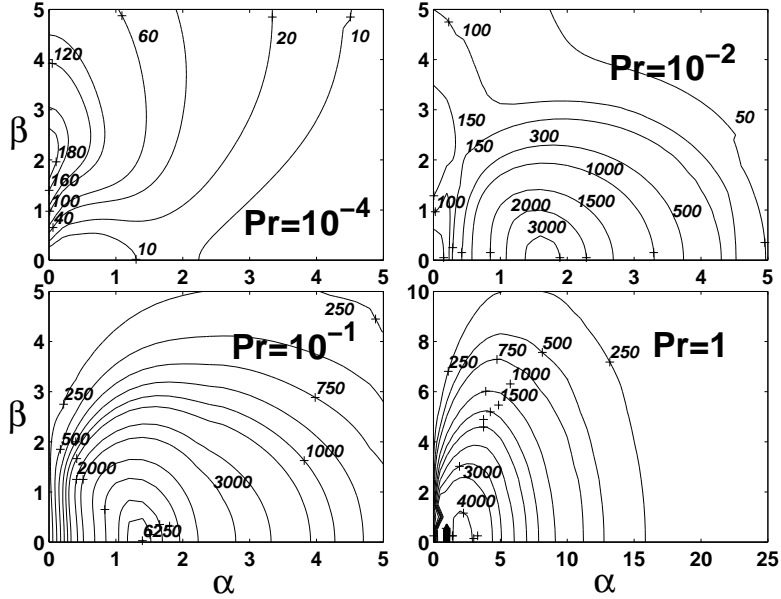}
\caption
{Asymmetric heating: contour plot of $G_{max}$ for  $T=25^o K$, 
$Re=1000$ for various Prandtl numbers. Note that for $Pr=1$ the scale 
employed is different.}
\label{pr1}
\mine
\end{figure}

It has been seen that the Prandtl number has a marginal effect 
on the most unstable linear mode. In contrast, we find here that
reducing heat diffusivity has a large destabilising effect on 
the transient growth of disturbance kinetic energy.
A dimensionless quantity for measuring growth is the energy norm 
defined as
\begin{eqnarray}
E=\int |u_p|^2+|v_p|^2+|w_p|^2+|\hat T|^2 dy.
\end{eqnarray}
There is some flexibility in defining the measure,
but the results are not expected to change qualitatively
[\cite{hanifi,bott2004}].
In figure \ref{pr1} for a temperature difference of $25^o K$ at a 
Reynolds number of $1000$ the effect of Prandtl number is shown.
As the Prandtl number is increased from $10^{-4}$ to $1$, 
the transient growth is seen to increase dramatically.
The large destabilisation comes from a new two-dimensional transient growth.
This is true for symmetric heating as well (not shown).
We now have a situation where transient growth dominates, but not
via the standard streamwise streaks and streamwise vortices.

\subsection{Effect of unstable density stratification}
\label{sec:buoy}
In their studies of stable thermal stratification \cite{bott2004} have found 
that as stratification increases flow becomes increasingly
stable, both in terms of exponential growth as well as transient growth. 
Viscosity variations were not accounted for in their calculation. 
In this paper, we concentrate on unstable thermal stratification.
\begin{figure}
\centering
\minp
\includegraphics[width=1.0\textwidth]{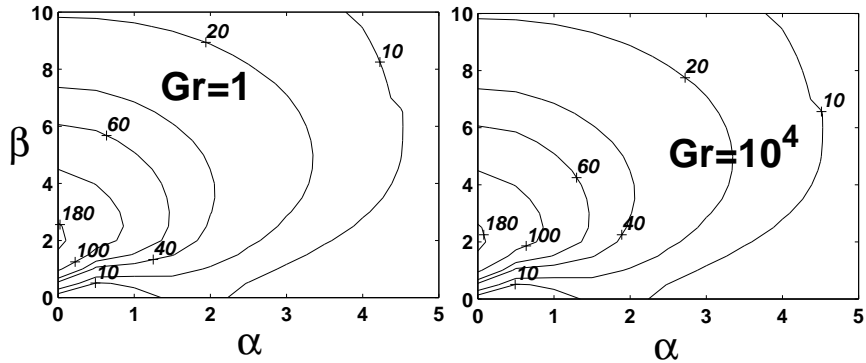}
\caption
{The transient growth at various Grashof number for $\Delta T=25^oK$, $Re=1000$ and $Pr=10^{-4}$. 
Since temperature perturbations diffuse away rapidly, buoyancy does not
have much effect.}
\label{grvspr4}
\mine
\end{figure}

\begin{figure}
\centering
\minp
\includegraphics[width=1.0\textwidth]{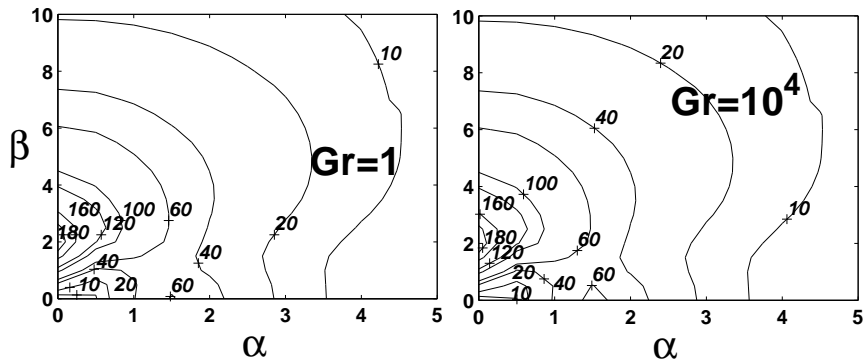}
\caption
{Same as figure \ref{grvspr4}, but for $Pr=10^{-3}$. The Grashof number has no 
effect upto a value of
$\sim 10^4$. At $Gr=10^4$ a new growth appears at $\beta=0$, which will
dominate at higher Prandtl number.}
\label{grvspr3}
\mine
\end{figure}

\begin{figure}
\centering
\minp
\includegraphics[width=1.0\textwidth]{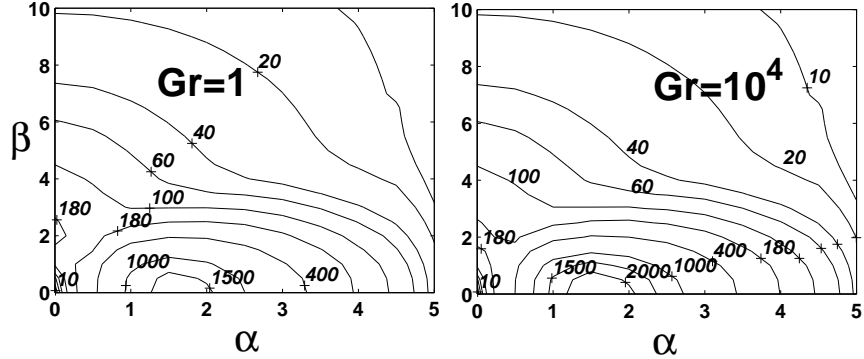}
\caption
{Same as figure \ref{grvspr4}, but for $Pr=10^{-2}$. The new subcritical
mode is now dominant at low Grashof number as well. The spanwise-independence
of the largest transient growth, making the Poiseuille-Rayleigh-Benard problem
essentially two-dimensional, is unusual.  }
\label{grvspr2}
\mine
\end{figure}

\begin{figure}
\centering
\minp
\includegraphics[width=1.0\textwidth]{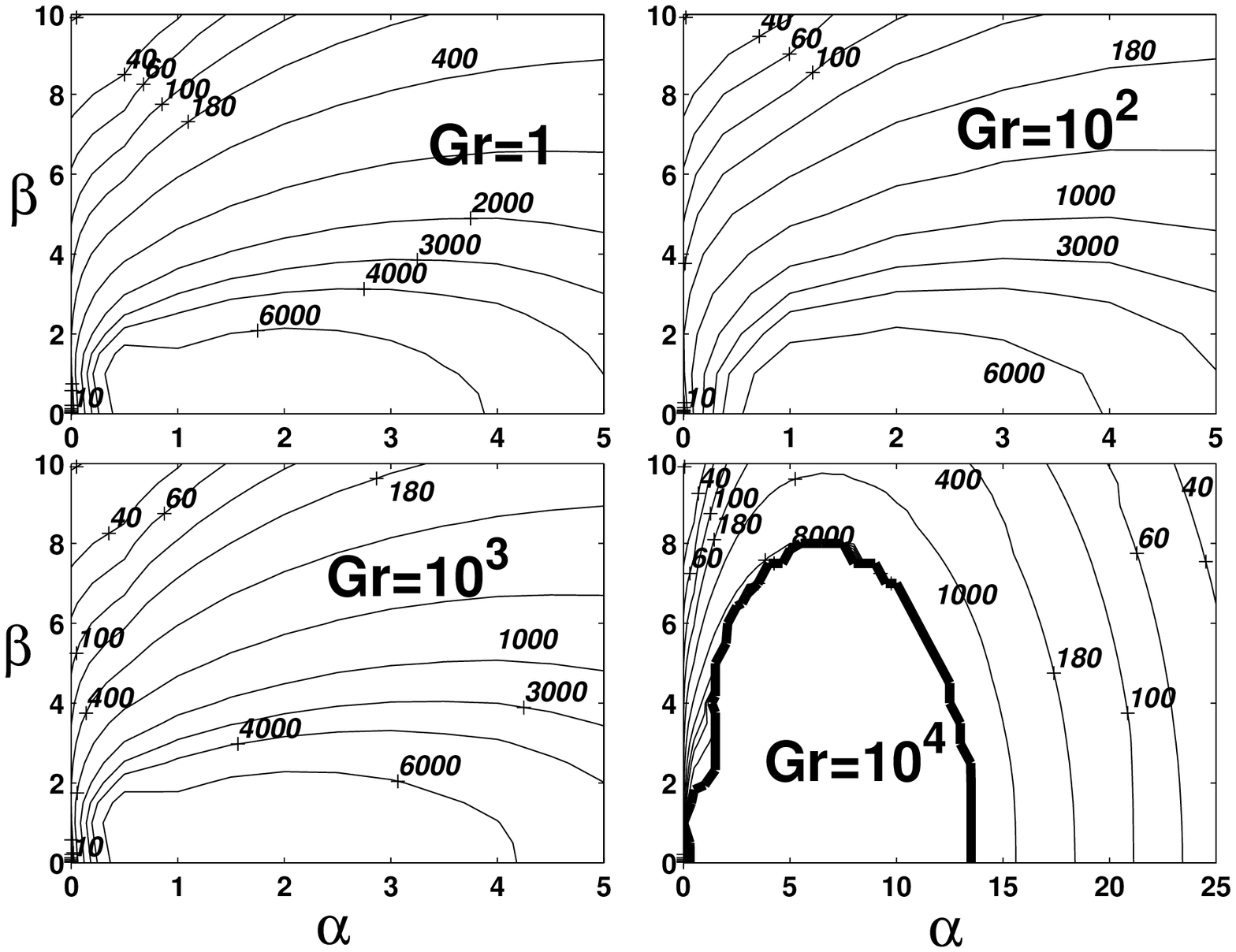}
\caption
{Same as figure \ref{grvspr4}, except that $Pr=1$. The transient growth
is even higher than before. The region inside the thick curve for $Gr=10^4$ 
is linearly unstable.}
\label{grvspr0}
\mine
\end{figure}

Figures \ref{grvspr4} to \ref{grvspr0} 
show contour plots of maximum growth of transient energy for various 
Grashof numbers at increasing Prandtl number for a temperature difference of 
$\Delta T=25^oK$. As expected, when heat diffusivity is high, buoyancy
has little effect. For $Pr \le 0.01$, the Prandtl number dictates the 
instability, and buoyancy has very little effect. At $Pr=1$, however, the
situation is completely different. For Grashof numbers of 1000 and below
we see extremely large levels of subcritical transient growth. This growth
is two-dimensional. Above
this Grashof number of course, a linearly unstable mode exists.
The transient growth in unheated
channel flow is well known to display itself as streamwise-independent
structures, like streaks and vortices. Our results indicate that such
structures will not be much in evidence in heated flows at realistic
Prandtl or Grashof numbers. Rather, a spanwise-independent growth occurs. 
Experimental and numerical verification of this kind of transient
growth could have interesting consequences for wall heating as a control
option.

%% file: secondary.tex
\section{Secondary Instability}
 
A flow containing linear modes (either growing or decaying) of sufficient
amplitude $A_p$ can become unstable to new secondary modes of instability. In 
unheated channel flow, secondary instability is considered unimportant, since
it may only play a role when external disturbance levels are extremely small.
We show here that viscosity variations can significantly destabilise
the secondary mode, thus making it more relevant to the transition process.
The Prandtl number and Grashof number are set equal to zero.

The approach here is as in \cite{her2} and \cite{her1}.
All flow variables are decomposed in the form 
\beq 
u(x,y,z,t)= U(y) +A_p\hat{u}_p(x,y,t) +A_s \hat{u}_s(x,y,z,t), 
\eeq 
where $\hat u_p$ is the linear instability mode
of the previous section. A subscript $p$ has been introduced in this section
to denote primary mode of instability. $A_s$ is the amplitude of the secondary. 
Note that since only the least stable linear modes are relevant here, it is
sufficient to consider two-dimensional primary modes, by Squire's theorem.
 
The secondary perturbation quantities are assumed to be of the form
\begin{eqnarray}
&&(\hat{u}_s,\hat{v}_s,\hat{w}_s)=\frac{1}{2}\Big[(u_+,v_+,w_+)(y,t)e^{i(\alpha_+ x+\beta_s z)}\nonumber\\
&&+(u_-,v_-,w_-)(y,t)e^{i(\alpha_-x-\beta_s z)}+{\rm c.c.} \Big],
\label{velsec}
\end{eqnarray}
where $\alpha_+$ and $\alpha_-$ are the wave numbers of the secondary
waves in the streamwise direction, $\beta_s$ is the wave number in the
spanwise direction. The direct interaction between primary waves is
assumed to be negligible.
For the flow under consideration, the growth/ decay rates are so
small that $d{A_p}/d{t}$ can be neglected during one period of
time, and the primary flow may be taken to be periodic.
Substituting these decompositions in the momentum equations,
eliminating pressure and neglecting non-linear terms in the secondary 
disturbance, we arrive at the secondary disturbance equations. 
On averaging these over $x,z$ and $t$, only the resonant modes survive, 
which are given by 
\beqa
\alpha_++\alpha_-=\alpha. 
\eeqa 
The cases of $\alpha_+= \alpha/2$ and $\alpha_+=\alpha$ are called the 
subharmonic and the fundamental modes respectively. Using continuity the
streamwise component of secondary disturbance velocity is eliminated and
we get the secondary perturbation equations:
\begin{eqnarray}
&&-D\frac{\partial v_+}{\partial t}+s\frac{\partial f_+}{\partial t}=-sAf_++(AD-i\alpha_+(DU))v_+
-A_p\Big[\frac{i\alpha_+^2}{2\alpha_-}u_pD+\frac{v_p\alpha_+D^2}{2\alpha_-}\nonumber \\
&&+\frac{i(Du_p)\alpha_+}{2}\Big]v_-^*
+\frac{A_p\alpha_+^2}{2}\Big[-v_pD+i\alpha_-u_p+\frac{i\beta_s^2}{\alpha_-}u_p+
\frac{\beta_s^2}{\alpha_+\alpha_-}v_pD\Big]f_-^*,
\label{abse}
\end{eqnarray}
\begin{eqnarray}
&&\frac{\partial v_+}{\partial t}-D\frac{\partial f_+}{\partial t}=-Av_++(AD+(DA))f_+
-\frac{A_p(\alpha+\alpha_-)}{2}\Big[\frac{v_p}{\alpha_-}D-iu_p\Big]v_-^*\nonumber\\
&&+{A_p \over 2}\Big[-i(\alpha+\alpha_-)u_pD
-{i\alpha_-(Du_p)}+v_p\Big(\frac{\alpha_+\beta_s^2}{\alpha_-}+D^2\Big)\Big]f_-^*,
\label{sec2}
\end{eqnarray}
where
$A=[i\alpha_+U+\mu s-\mu d^2-\mu^\prime D]$,
$f_+=-\frac{i}{\beta_s}w_+$,
$D={d}/{dy}$.
Equations (\ref{abse}) \& (\ref{sec2}) and complementary
equations  for $v_-^*$ and $f_-^*$ are solved using a Chebyshev collocation
spectral method, with the boundary conditions $\hat{u}_s,\hat{v}_s,\hat{w}_s=0$ at $y=\pm1$.
The dispersion relation is $ F(A_p,\beta_s,m,Re,\alpha,c,)=0$ [see \cite{her2}].
The growth rate is highly sensitive to the primary amplitude level $A_p$, and
increases with increasing $A_p$.
The present computations are validated by comparing with the unstratified case
in \cite{her2} as discussed in \cite{phd}.

\subsection{Asymmetric heating}

\begin{figure}
\centering
\minp
\includegraphics[width=1.0\textwidth]{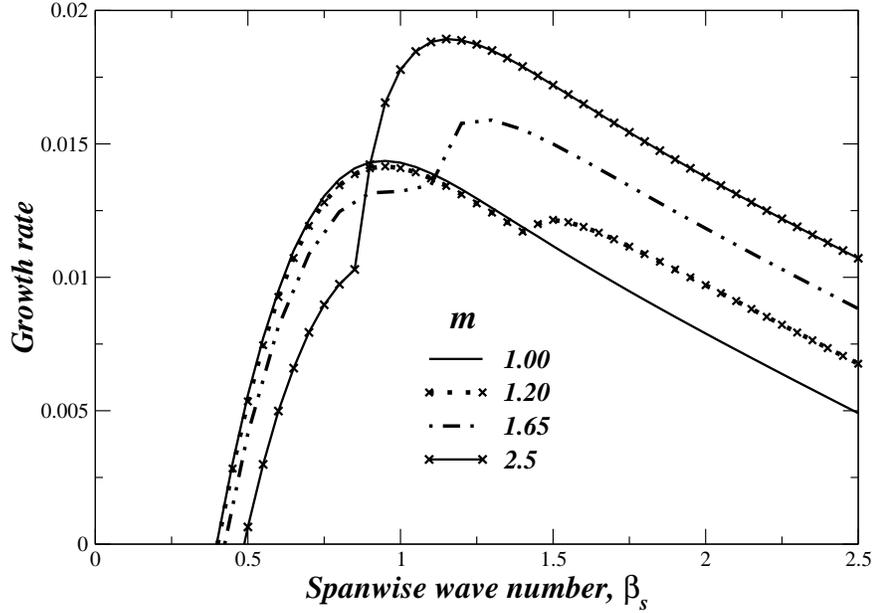}
\caption {Asymmetric heating: dependence of growth rate on spanwise wave number of
the secondary disturbance for various viscosity ratios,
subharmonic case.
 $\alpha=1.0$, $A_p=0.01$, $Re=5000$.}
\label{growth}
\mine
\end{figure}

A value of $A_p=0.01$ is
taken, to be representative of an intermediate level of primary disturbance. 
The variation of secondary growth rate $\omega_{is}$ with the spanwise wave number 
for various viscosity ratios is plotted in figures \ref{growth}. As the
temperature difference increases, a second highly unstable mode appears.
This mode is three-dimensional, the spanwise and streamwise
wavelengths being very close to each other. The modes which are closer to
two-dimensional are now stabilised. A nonlinear or transient growth triggered by
this new mode could mean that transition to turbulence proceeds
somewhat differently, but further studies are needed to evaluate this.

\begin{figure}
\centering
\minp
\includegraphics[width=1.0\textwidth]{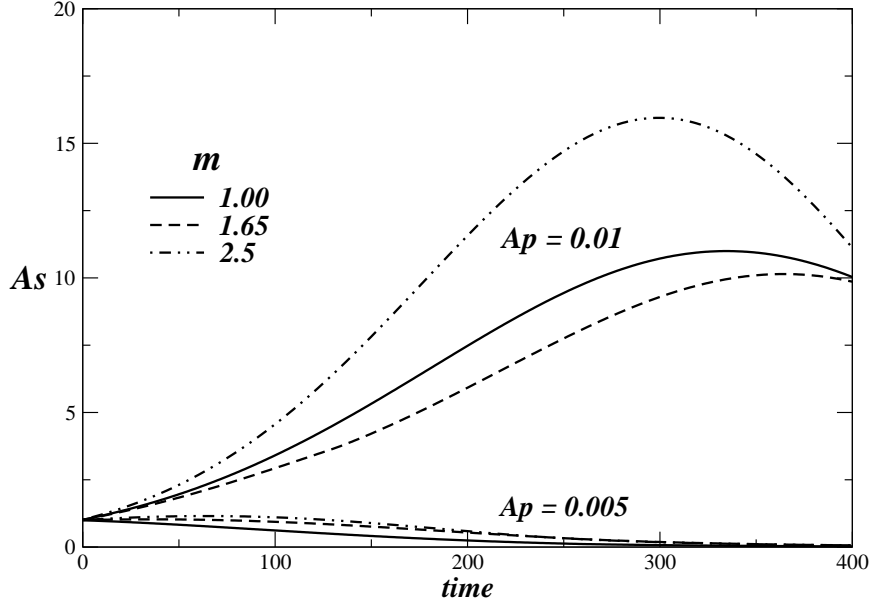}
\caption
{Asymmetric heating: variation with time of the amplitude of secondary 
disturbance for
two sets of initial $A_p$. $\alpha=1.0$, $\beta_s=1.0$, $Re=5000$, Subharmonic mode.}
\label{Asvst3}
\mine
\end{figure}

While we have taken the amplitude of the primary mode to be constant, it is in 
fact, at these Reynolds numbers, a known slowly decaying function of time.
Integrating instantaneous results over long times, we can compute the time 
dependence of amplitude of the secondary mode. This approach is a counterpart 
in time of the assumption of parallel flow in flows which vary slowly in $x$.
The amplitude of the subharmonic secondary mode $A_s$ is shown as a function of time
in figure \ref{Asvst3}. At low initial $A_p$, secondary modes are always stable 
while for higher $A_p$ significant growth is displayed up to large times.
At low Reynolds number, the initial $A_p$ needed for a sustained secondary 
instability growth is very high. 

\begin{figure}
\centering
\minp
\includegraphics[width=1.0\textwidth]{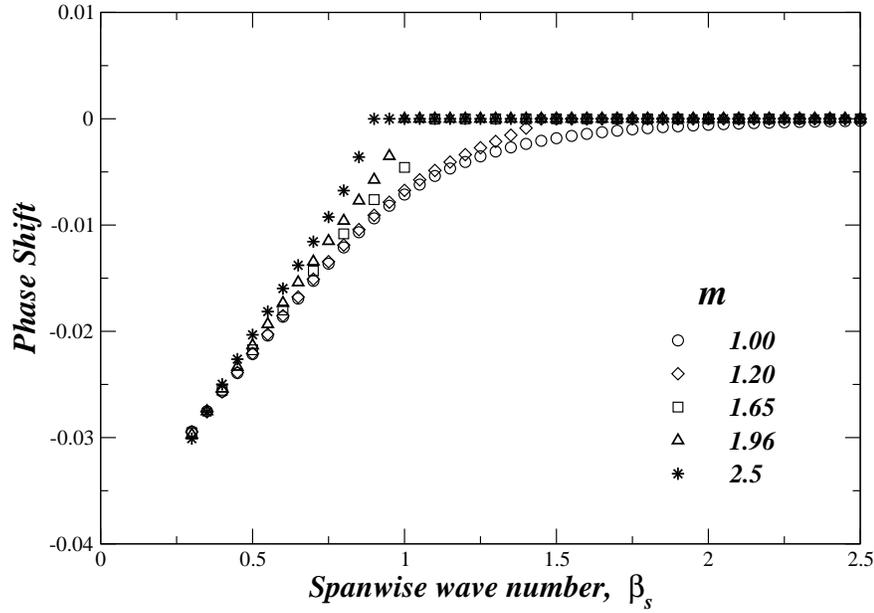}
\caption
{Asymmetric heating: variation of $P_S$ with spanwise wave number for various viscosity
 ratios, subharmonic case. $\alpha=1.0$, $A_p=0.01$, $Re=5000$.}
\label{freq}
\mine
\end{figure}

In figure \ref{freq} the phase shift $P_S=\omega_p\frac{\alpha_+}{\alpha}-\omega_s$
is shown as a function of the spanwise wave number. 
The phase locking of the subharmonic wave (i.e., where $P_S$ is zero)
is achieved at an earlier $\beta_s$ than for the
unstratified case.

\subsection{Symmetrically heated channel}
\begin{figure}
\centering
\minp
\includegraphics[width=1.0\textwidth]{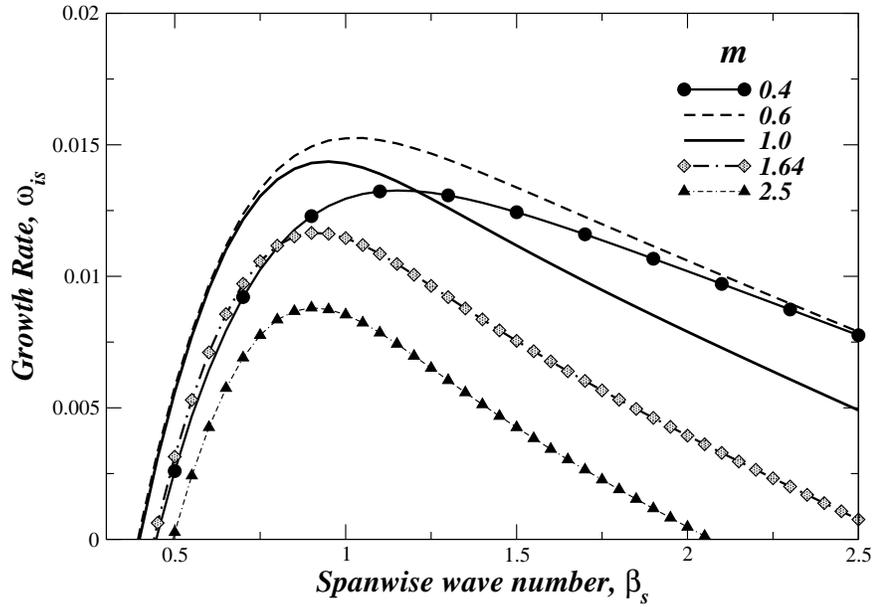}
\caption {The dependence of growth rate on spanwise wave number of
the secondary disturbance for various viscosity ratios,
subharmonic case.
 $\alpha=1.0$, $A_p=0.01$, $Re=5000$, symmetric heating.}
\label{grvsb_re5000sym}
\mine
\end{figure}

Figure \ref{grvsb_re5000sym} shows the
secondary perturbation growth rate variation with spanwise wave
number $\beta_s$. A stabilisation with increase in viscosity ratio, especially 
when $m>1$, is evident. This behaviour is remarkable, being counter intuitive and
 opposite to the behaviour of the primary instability mode.
The phase locking behaviour is similar to the asymmetrically heated case.

\begin{figure}
\centering
\minp
\includegraphics[width=1.0\textwidth]{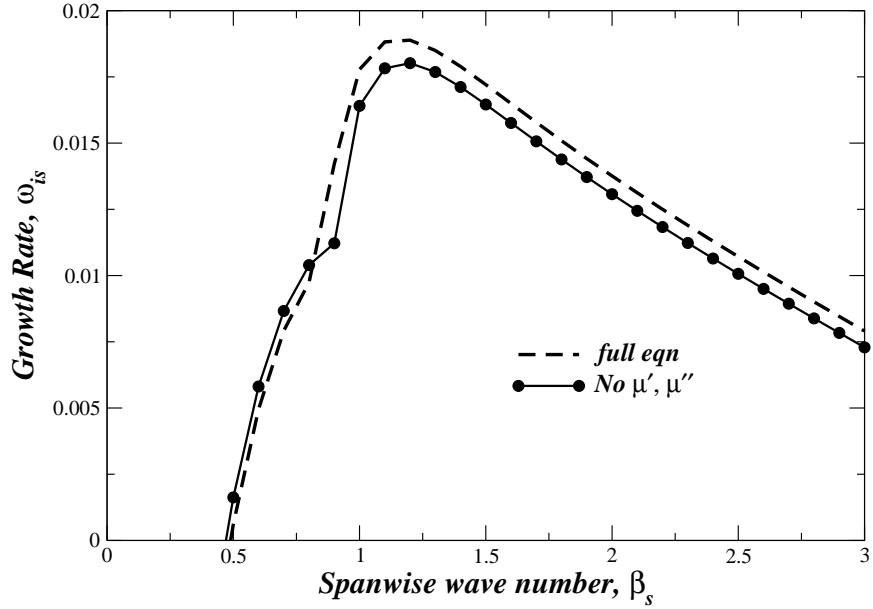}
\caption{The new mode of instability arises from changes in the
velocity profile. This is verified by switching off the viscosity gradient 
terms in the secondary stability equation. The dashed lines are for the full 
equation and the filled circles are with the viscosity gradient set to 
zero in the instability calculations, but retained in the mean flow
computations. $\Delta T=50 (m=2.5)$, 
 $\alpha=1.0$, $\alpha_s=0.5$, $A_p=0.01$, $Re=5000$, asymmetric heating.}
\label{invis_sec}
\mine
\end{figure}

We have seen that in linear disturbance growth, the mean velocity profile
(via the $U^{\prime\prime}$ term) has a dominant role.
In the case of secondary growth as well, \cite{orszag83} have argued that
inviscid effects are dominant, and act through vortex stretching and tilting.
We are not able to make a conclusive statement on this, but it seems that
heating affects secondary instability by an inviscid mechanism, through 
changes in the velocity profile. This is demonstrated in figure
\ref{invis_sec}, where it is seen that switching off the viscosity 
gradient terms makes little difference to the result.
Also, it is not evident why the sign of instability is opposite to that of
primary modal growth.
The relevant inference here is that viscosity stratification alone can have unexpected
effects on the various mechanisms leading to transition to turbulence. 

%% file: conc.tex
\section{Conclusion}
The control of the flow using heating or cooling of the surface has 
long been practised, especially in open flows. In plane channel flow
we have conducted a comprehensive study of the effect of heat,
and show that there is no unique direction (either towards or away
from stabilisation) in which the flow responds. Linear
stability results are consistent with earlier studies in that
the most unstable linear mode is suppressed when viscosity decreases
towards the wall. Also the effect of Prandtl number is negligible. 
For Grashof number between about 3000 and 25000, separate modes of 
Rayleigh-Benard and
Tollmien-Schlichting instability are evident, the former is at low
Reynolds number. At higher levels of buoyancy, the modes merge.

The transient growth of disturbances is unaffected by viscosity
stratification, but hugely increased by reduced heat diffusivity.
Both of these are counter to the effect on the least stable
linear mode. The Prandtl number is thus not an unimportant 
parameter, as was hitherto assumed. Transient growth is also 
very high in the presence of buoyancy of the appropriate sign. 
With increasing Prandtl and/or Grashof number, the growth is 
two-dimensional, not in streamwise streaks, which is quite 
unusual for transient growth. 

Secondary instability of Tollmien-Schlichting waves is not
considered an important player in the transition to turbulence
in unheated channel flow, unless the free stream is unrealistically
quiet. We show that a new destabilising mode appears with heating,
for the case where viscosity is decreasing towards the wall!

It is hoped that this work with give impetus to experimental
and computational studies to check these predictions and to
explore wall heating in all its aspects as a control strategy 
for channel and pipe flows.

\begin{acknowledgments}
We are indebted to Prof. O.N. Ramesh for discussions throughout the 
course of this work and for reading the manuscript. 
A part of the work was carried out during the tenure of AS as a 
PhD student at the Indian Institute of Science. 
RG would like to thank Defence R \& D Organisation, India for 
funding the project.
\end{acknowledgments}

\bibliographystyle{jfm}
\bibliography{myref_thesis}